\documentclass{camera}
\usepackage{graphicx}  

\begin{document}

%
\title{Projectile deformation effects in the breakup of $^{37}$Mg}

%
\author{Shubhchintak$^{1}$, R. Chatterjee$^2$  \and R.Shyam$^{3,2}$}

%
\organization{{\em $^1$Department of Physics and Astronomy, Texas A$\&$M University - Commerce, Tx-75428, USA}\\
{\em $^2$Department of Physics, Indian Institute of Technology - Roorkee, 247667, India } \\
{\em $^3$ Theory Group, Saha Institute of Nuclear Physics, 1/AF Bidhannagar,
Kolkata 700064, India}\\}

\maketitle

\begin{abstract}
We study the breakup of $^{37}$Mg on Pb at 244MeV/u 
with the recently developed extended theory of 
Coulomb breakup within the post-form finite range distorted wave Born approximation 
that includes deformation of the projectile. Comparing our calculated cross section with the available Coulomb breakup data  we determine the possible ground state 
configuration of $^{37}$Mg.
\end{abstract}

%
\section*{Introduction and Formalism}
Coulomb breakup of nuclei away from the valley of stability has been one of the most successful probes to unravel their structure. However, it is only recently that one is venturing into medium mass nuclei like $^{23}$O \cite{rc1} and $^{31}$Ne \cite{shubh}, especially in and around the so called ``island of inversion". This is a very new and exciting development which has expanded the field of light exotic nuclei to the deformed medium mass region.

We consider the elastic breakup of a two body composite `deformed' projectile $a$ 
in the Coulomb field of a target $t$, $a+t\rightarrow b+c+t$, where projectile $a$ 
breaks up into fragments $b$ (charged) and $c$ (uncharged). The reduced transition 
amplitude, $\beta_{\ell m}$, for this reaction is given by \cite{rc0}
\begin{eqnarray}
\beta_{\ell m} &=& \left\langle e^{i(\gamma{\bf q_c} - \alpha{\bf K}).{\bf r_1}}
\left|V_{bc}({\bf{r}}_{1})\right|\phi_{a}^{\ell m}({\bf r}_1)\right\rangle 
\nonumber\\
&\times&\left\langle \chi_{b}^{(-)}({\bf q_b},{\bf r_i})e^{i\delta 
{\bf q_c}.{\bf r_i}}|\chi_{a}^{(+)}({\bf q}_a,{\bf r_i})\right\rangle. \label{a4.2}
\end{eqnarray}

The ground state wave function of the projectile $\phi_{a}^{lm}({\bf r}_1)$ appears 
in the first term (vertex function), while the second term that describes the 
dynamics of the reaction, contains the Coulomb distorted waves $\chi^{(\pm)}$. 
This can be expressed in terms of the bremsstrahlung integral. $\alpha$, $\gamma$ 
and $\delta$ are the mass factors pertaining to the three-body Jacobi coordinate 
system (see Fig. 1 of Ref. ~\cite{shubh}). In Eq. 1, ${\bf K}$ is an effective local 
momentum appropriate to the core-target relative system and ${\bf q}_i$'s ($i = a, 
b, c$) are the Jacobi wave vectors of the respective particles. 

$V_{bc}({\bf{r}}_{1})$ [in Eq. (\ref{a4.2})] is the interaction between $b$ and $c$, 
in the initial channel. We introduce an axially symmetric quadrupole-deformed 
potential, as
\begin{eqnarray}
V_{bc}({\bf r}_1) &=& \frac{V_{ws}}{1+exp(\frac{r_1-R}{a})} \nonumber\\ 
&~& -\beta_2 RV_{ws} \frac{df(r_1)}{dr_1} Y^{0}_{2}(\hat {\bf r}_1), \label{a4.3}
\end{eqnarray}
where $V_{ws}$ is the depth of the spherical Woods-Saxon potential, $\beta_2$ is the 
quadrupole deformation parameter. The first part of the Eq. (\ref{a4.3}) is the 
spherical Woods-Saxon potential $V_s(r_1)$ with radius $R = r_0A^{1/3}$. $r_0$ and 
$a$ being the radius and diffuseness parameters, respectively. To preserve the 
analyticity of our method, we calculate the radial part of the ground state wave 
function of the projectile using the undeformed Woods-Saxon potential (radius and 
diffuseness parameters taken as 1.24 fm and 0.62 fm respectively, which reproduce 
the ground state binding energy). We emphasize that the deformation parameter 
($\beta_2$) has already entered into the theory via $V_{bc}$ in Eq. (\ref{a4.2}).  
For more details on the formalism we refer to Ref. \cite{shubh2}.

\section*{Results and discussions}
The nucleus $^{37}$Mg has a large uncertainty in its one-neutron separation energy
($0.162\pm0.686$ MeV \cite{wan12}) and has controversies regarding its ground state 
spin-parity. Recently measured large breakup cross section \cite{kob14} and reaction 
cross section \cite{takechi2} seems to suggest a halo structure in $^{37}$Mg. 

The nuclei in island of inversion are expected to have significant components of 
$2p-2h$ [$\nu(sd)^{-2}(fp)^2]$ neutron intruder configurations. Indeed, in 
Ref.~\cite{kob14}, it has been argued that the valence neutron in $^{37}$Mg$_{gs}$ 
is most likely to have a spin parity ($J^\pi$) of $3/2^-$ that corresponds to the 
$2p_{3/2}$ orbital. For the sake of completeness, in this work we have considered neutron removal from 
$2p_{3/2}$, $2s_{1/2}$ and $1f_{7/2}$ orbitals. 

\begin{figure}
\centering
\includegraphics[height=5cm, clip,width=0.5\textwidth]{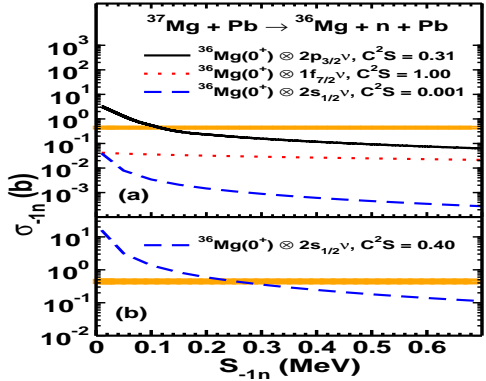}
\caption{\label{fig2} (a) Pure Coulomb total one-neutron removal cross section, 
$\sigma_{-1n}$, in the breakup reaction of $^{37}$Mg on a Pb target at 
244 MeV/nucleon beam energy as a function of one-neutron separation energy 
S$_{-1n}$ obtained with configurations $^{36}$Mg$(0^+)\otimes 2p_{3/2}\nu$ 
(solid line),  $^{36}$Mg$(0^+)\otimes 2s_{1/2}\nu$ (dashed line) and 
$^{36}$Mg$(0^+)\otimes 1f_{7/2}\nu$ (dotted line) for $^{37}$Mg$_{gs}$ using the 
shell model spectroscopic factors ($C^2S$) as indicated in each case. The 
experimental cross section (taken from Ref.~\cite{kob14}) is shown by the shaded 
band. (b) Same reaction as in (a) for the $^{36}$Mg$(0^+)\otimes 2s_{1/2}\nu$ 
configuration using the $C^2S$ value of 0.40 as deduced in Ref.~\cite{kob14}.   
}
\end{figure}

In Fig.~1(a), we show the results of our calculations for the pure Coulomb 
$\sigma_{-1n}$ in the breakup reaction of $^{37}$Mg on a Pb target at the beam 
energy of 244 MeV/nucleon as a function of $S_{-1n}$ corresponding to one-neutron 
removal from the $2p_{3/2}$, $2s_{1/2}$ and $1f_{7/2}$ orbitals. For $C^2S$ we 
have used the shell model values as given in Ref.~\cite{kob14}, which are 0.31 
and 0.001 for the $^{36}$Mg$(0^+)\otimes 2p_{3/2}\nu$, and 
$^{36}$Mg$(0^+)\otimes 2s_{1/2}\nu$ configurations, respectively. However, for the 
$^{36}$Mg$(0^+)\otimes 1f_{7/2}\nu$ configuration the SM $C^2S$ is not given in 
this reference. Therefore, we have assumed a $C^2S$ of 1.0 for this case. The 
shaded band in this figure shows the corresponding measured cross section taken 
from Ref.~\cite{kob14} with its width representing the experimental uncertainty. 
We note that calculated cross sections obtained with the $^{36}$Mg$(0^+)\otimes 
2p_{3/2}\nu$ configuration (solid line in Fig.~1), overlap with the experimental 
band in the $S_{-1n}$ region of $0.10 \pm 0.02$. Theoretical cross sections for the 
$2p_{1/2}$ case are almost identical to those of the $2p_{3/2}$ case. On the other 
hand, for the $^{36}$Mg$(0^+)\otimes 2s_{1/2}\nu$ and $^{36}$Mg$(0^+)\otimes 
1f_{7/2}\nu$ configurations there is no overlap between calculated cross sections 
and the data band. Therefore, our results support a $J^\pi$ = 3/2$^-$ ground state 
for $^{37}$Mg with a one-neutron separation energy of $0.10 \pm 0.02$. The $S_{-1n}$ 
deduced in our work is closer to the evaluated value of $0.16 \pm 0.68$~\cite{wan12}, 
with lesser uncertainty.   

Nevertheless, with $C^2S$ for the $^{36}$Mg$(0^+)\otimes 2s_{1/2}\nu$ configuration 
as extracted in Ref.~\cite{kob14} (0.40), the cross section curve for this case 
will also overlap with the experimental data band as shown in Fig.~1(b). However, 
the extracted $S_{-1n}$ in our case is $0.26 \pm 0.04$ MeV, instead of
0.40$^{+0.19}_{-0.13}$ MeV deduced in Ref.~\cite{kob14}. In fact, with the combination 
of the mean values of $S_{-1n}$ and $C^2S$ (0.40 MeV and 0.40) for this configuration 
deduced in Ref.~\cite{kob14}, there would be no overlap between the theoretical 
cross section and the experimental data band in our calculations.  

\begin{figure}
\centering
\includegraphics[height=4cm, clip,width=0.5\textwidth]{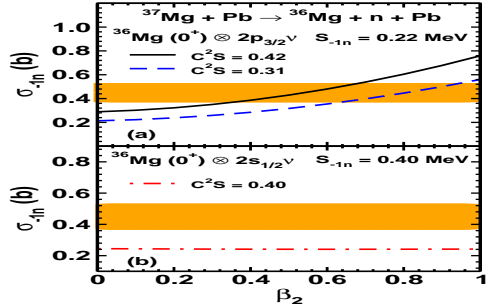}
\caption{\label{fig4} (a) $\sigma_{-1n}$ as a function of the deformation parameter
$\beta_2$ in the Coulomb breakup of $^{37}$Mg on a Pb target at the beam energy of
244 MeV/nucleon with the configuration $^{36}$Mg$(0^+)\otimes 2p_{3/2}\nu$ for 
$^{37}$Mg$_{gs}$. The S$_{-1n}$ is taken to be 0.22 MeV with $C^2S$ values being 
0.42 (solid line) and 0.31 (dashed line). (b) Same as in Fig.~2(a) for 
$^{36}$Mg$(0^+)\otimes 2s_{1/2}\nu$ configuration with $C^2S$ and $S_{-1n}$ of 0.40
and 0.40 MeV, respectively. In both (a) and (b) the experimental data  (shown by the 
shaded region) are taken from Ref. \cite{kob14}.}
\end{figure}

In Fig.~2(a), we show our results for $\sigma_{-1n}$ as a function of $\beta_2$ 
for the $^{36}$Mg$(0^+)\otimes 2p_{3/2}\nu$ configuration of $^{36}$Mg$_{gs}$ 
with $C^2S$ values of 0.42 and 0.31 and taking a $S_{-1n}$ of 0.22 MeV ( as in Ref. \cite{kob14}) in both 
the cases. For $\beta_2=0$, the $\sigma_{-1n}$ in each case is the same as that 
shown in Fig.~1, which is below the experimental data. With increasing $\beta_2$, 
the cross sections in both the cases increase, and the overlaps between 
calculations and the data take place in ranges $0.35 <\beta_2<0.68$ and  
$0.62<\beta_2<0.94$ for $C^2S$ of 0.42 and 0.31, respectively. We add that if for 
$C^2S$ = 0.31 the $S_{-1n}$  as deduced in Fig.~1 is taken then calculated cross 
section would overlap with the data band at much lower values of $\beta_2$. 

In Fig.~2(b) we show the same results for the $^{36}$Mg$(0^+)\otimes 2s_{1/2}\nu$
configuration with $C^2S$ and $S_{-1n}$ values of 0.40 and 0.40 MeV, respectively, 
(which is the mean value of these quantities as deduced in Ref.~\cite{kob14}). In 
this case, we see that  in contrast to the results in Fig.~2(a), there is no overlap 
between calculated cross sections and the data band for any value of $\beta_2$ in 
the range of 0$-$1.0. We have checked that the situation remains the same for 
$\beta_2$ $>$ 1.0. Moreover, the contribution of the deformation term to the 
cross section is substantially low for the $s$-wave configuration, which results in
almost constant $\sigma_{-1n}$ as a function of $\beta_2$ as seen in Fig.~2(b). 
Therefore, in our calculation, even with deformation a $s$-wave configuration is 
ruled out for $^{37}$Mg$_{gs}$ for the $C^2S$ and $S_{-1n}$ combination deduced in 
Ref.~\cite{kob14}.

\section*{Conclusion}
In this paper we have studied the Coulomb breakup reaction $^{37}$Mg + Pb $\to$ 
$^{36}$Mg + n + Pb at the beam energy of 244 MeV/nucleon, within the framework of 
the post form finite range distorted wave Born approximation theory that is extended 
to include the projectile deformation effects. In this formalism the transition 
amplitude is factorized into two parts - one containing the dynamics of the reaction 
and the another the projectile structure informations such as the fragment-fragment 
interaction and the corresponding wave function in its ground state.  Analytic 
expressions can be written for both  parts. This formalism opens up a route to 
perform realistic quantum mechanical calculations for the breakup of neutron-drip 
line nuclei in the medium mass region that can be deformed.

We calculated the total one-neutron removal cross sections ($\sigma_{-1n}$) in 
this reaction and compared our results with the corresponding data reported in a 
recent publication~\cite{kob14}. Our calculations seem to favor a $J^\pi$ = 3/2$^-$ 
spin assignment to the $^{37}$Mg ground state with one-neutron separation energy 
($S_{-1n}$) of $0.10 \pm 0.02$ MeV, if the spectroscopic factor ($C^2S$) for this state 
is taken to be the corresponding shell model value of 0.31. However, the deduced 
$S_{-1n}$ depends on the chosen value of $C^2S$. Our study shows that $S_{-1n}$ rises 
steadily with increasing $C^2S$. Indeed, due to the uncertainty in the $C^2S$ value 
for the $^{36}$Mg($0^+$) $\otimes$ ${2s_{1/2}}\nu$ configuration for the $^{37}$Mg 
ground state, the $J^\pi$ = 1/2$^+$ spin assignment to it can not be fully excluded
based on the present data.

In order to gain more insight in the ground state structure of $^{37}$Mg, we studied 
the effect of the projectile deformation on $\sigma_{-1n}$. We find that for the 
configuration $^{36}$Mg($0^+$) $\otimes$ ${2p_{3/2}}\nu$ for the $^{37}$Mg ground 
state, the calculated $\sigma_{-1n}$ overlaps with the experimental data band in 
certain range (that depends on the value of $C^2S$) of the quadrupole deformation 
parameter ($\beta_2$). However, with the $^{36}$Mg($0^+$) $\otimes$ ${2s_{1/2}}\nu$ 
configuration, the overlap does not occur between calculated and measured 
$\sigma_{-1n}$ for  any reasonable combination of $\beta_2$ and $C^2S$ values. This 
supports the $J^\pi$ = 3/2$^-$ spin assignment for the $^{37}$Mg ground state.

However, for unambiguous confirmation one also needs to calculate \cite{shubh2} more exclusive observables such as the core-valence 
neutron relative energy spectra, the energy-angle and the angular distributions of 
the emitted neutron and the parallel momentum distribution of the core fragment. 
The position of the peak as well as the magnitude of the cross section near the peak
of the core-valence neutron relative energy spectra would to be dependent on the 
configuration of the projectile ground state as well as on its deformation. 

 Our study is expected to provide motivation for future 
experiments on breakup reactions of the neutron rich medium mass nuclei.

\section*{Acknowledgments}
This work is supported by the Council of Scientific and Industrial Research (CSIR) 
and the Department of Science and Technology(SR/S2/HEP-040/2012), Govt. Of India . 

%

\end{document}